\documentclass[oldversion]{aa}  

\newcommand{\lens}{MACS\,J0717}

\usepackage{graphicx}
\usepackage{txfonts}
\usepackage{natbib}
\begin{document}
 
\title{
Strong-lensing analysis of MACS\,J0717.5+3745 from \\ \emph{Hubble Frontier Fields} observations: \\ How well can the mass distribution be constrained?}
   \titlerunning{MACS\,J0717}
   \authorrunning{Limousin et~al.}
   %\subtitle{The Core of MACS\,J0717.5+3745 at $z=0.55$}
   \author{M. Limousin\inst{1},
J. Richard\inst{2},
E. Jullo\inst{1}
M. Jauzac\inst{3,4,5},
H. Ebeling\inst{6},
M. Bonamigo\inst{1},
A. Alavi\inst{7},
B. Cl\'ement\inst{2},
C. Giocoli\inst{1},
J.-P. Kneib\inst{8,1},
T. Verdugo\inst{9,10},
P. Natarajan\inst{11},
B. Siana\inst{7},
H. Atek\inst{8,11}
\& M. Rexroth\inst{8}
      \thanks{Based on observations obtained with the \emph{Hubble Space Telescope}}
       }
   \offprints{marceau.limousin@lam.fr}

   \institute{
        Laboratoire d'Astrophysique de Marseille, UMR\,6610, CNRS-Universit\'e de Provence,
        38 rue Fr\'ed\'eric Joliot-Curie, 13\,388 Marseille Cedex 13, France
        \and
        CRAL, Observatoire de Lyon, Universit\'e Lyon 1, 9 Avenue Ch. Andr\'e, 69561 Saint Genis Laval Cedex, France
        \and
        Centre for Extragalactic Astronomy, Department of Physics, Durham University, Durham DH1 3LE, U.K.
        \and
        Institute for Computational Cosmology, Durham University, South Road, Durham DH1 3LE, U.K.
        \and
        Astrophysics and Cosmology Research Unit, School of Mathematical Sciences, University of KwaZulu-Natal, Durban 4041, South Africa
        \and
        Institute for Astronomy, University of Hawaii, 2680 Woodlawn Dr, Honolulu, HI 96822, U.S.A.
        \and
        Department of Physics and Astronomy, University of California Riverside, CA 92521, U.S.A.
        \and
        Laboratoire d'Astrophysique, Ecole Polytechnique F\'ed\'erale de Lausanne, Observatoire de 
Sauverny, CH-1290 Versoix, Switzerland
	\and
	Universidad Nacional Aut\'onoma de M\'exico, Instituto de Astronom\'ia, Apdo.Postal 106, Ensenada, B.C. 22860 M\'exico
	\and
	Instituto de F\'{\i}sica y Astronom\'{\i}a, Universidad de Valpara\'{\i}so,  Avenida Gran Breta\~{n}a 1111, Valpara\'{\i}so, Chile
        \and
        Department of Astronomy, Yale University, 260 Whitney Avenue, New Haven, CT 06511, U.S.A.
              }

   %\date{Received...}
   
  \abstract
   {
We present a strong-lensing analysis of MACSJ0717.5+3745 (hereafter \lens), based on the full depth of the \emph{Hubble Frontier Field} (HFF) observations, which brings the number of multiply imaged systems to 61, ten of which have been spectroscopically confirmed. The total number of images comprised in these systems rises to 165, compared to 48 images in 16 systems before the HFF observations. Our analysis uses a parametric mass reconstruction technique, as implemented in the \textsc{Lenstool} software, and the subset of the 132 most secure multiple images to constrain a mass distribution composed of four large-scale mass components (spatially aligned with the four main light concentrations) and a multitude of galaxy-scale perturbers. We find a superposition of \emph{cored} isothermal mass components to provide a good fit to the observational constraints, resulting in a very shallow mass distribution for the smooth (large-scale) component. Given the implications of such a \emph{flat} mass profile, we investigate whether a model composed of "peaky" non-cored mass components can also reproduce the observational constraints.
We find that such a \emph{non-cored} mass model reproduces the observational constraints equally well,
in the sense that both models give comparable total RMS.
Although the total (smooth dark matter component plus galaxy-scale perturbers) mass distributions of both models are consistent, as are the integrated two-dimensional mass profiles, 
we find that the smooth and the galaxy-scale components are very different. 
We conclude that, even in the HFF era, the generic degeneracy between smooth and galaxy-scale components is not broken, in particular in such a complex galaxy cluster. 
Consequently, \emph{insights into the mass distribution of \lens\ remain limited}, emphasizing the need for additional probes beyond strong lensing.

Our findings also have implications for estimates of the lensing magnification. We show that the amplification difference between the two models is larger than the error associated with either model, and that this additional systematic uncertainty is approximately the difference in magnification obtained by the different groups of modelers using pre-HFF data. This uncertainty decreases the area of the image plane where we can reliably study the high-redshift Universe by 50 to 70\%.
   }


   \maketitle
%________________________________________________________________

\section{\lens\ in the \emph{Hubble Frontier Field era}}

\lens, a galaxy cluster located at $z=0.55$, is well established as one of the most massive and complex merging structures known so far, based on extensive optical \citep{harald0717,highzmacs,galax0717,morphodensity0717}, radio \citep{edge0717,vanweeren0717,bonafede0717,cral0717}, X-ray \citep{ma0717xray}, and Sunyaev Zel'dovich studies \citep{laroque03,mroczkowski0717,sayers0717}. Acting as a powerful gravitational lens, the system has been investigated both in the strong- \citep{adi0717,my0717,richard14,diego0717} and in the weak-lensing regimes \citep{jauzac0717,elinor0717}, further underlining its position as one of the most complex, dynamically active, and massive clusters studied to date. 

As a result, \lens\ has been selected as a target by the CLASH program \citep{clash} and more recently by the Hubble Frontier Field (HFF) initiative. The HFF project constitutes the largest commitment of Hubble Space Telescope (HST) time ever made to the exploration of the high-redshift Universe via gravitational lensing by massive galaxy clusters: 140 orbits of HST time have been devoted to deep imaging observations of six galaxy clusters. Each cluster is observed for 20 orbits in each of three ACS filters (F435W, F606W and F814W), and in each of four WFC3 passbands (F105W, F125W, F140W and F160W). More information on the HFF initiative can be found on their  website\footnote{http://www.stsci.edu/hst/campaigns/frontier-fields/}.
%These deep observations allow to detect many new multiply imaged systems compared to pre-HFF observations. 
%Concerning the first two clusters observed, Abell\,2744 and  MACS\,J0416, we have been able to
%detect up to 150-200 multiple images per cluster, compared to about 40-50 before the HFF observations 
%\citep{0416hff,a2744hff}.

In this paper, we use the full set of HFF observations in order to pursue a strong-lensing analysis of \lens. Our aims are twofold: (i) First, to study the mass distribution in detail, which entails estimating the total projected mass of the core of \lens, quantifying the location and shape of the mass components relative to the gas and galaxy distribution, and using the results to gain further insight into the ongoing merging processes in \lens, which may have broader implications for our understanding of structure formation and evolution. (ii) Second, to provide the community with a calibrated mass model of \lens, thereby enabling its use as an efficient gravitational telescope for studies of the high-redshift Universe, which is the primary scientific goal of the HFF program.

Previous analyses revealed that \lens\ is undergoing multiple merger events, reflected in its quadri-modal mass distribution, with a filament extending to the south-east. Using X-ray and optical data (both imaging and spectroscopy), \citet{ma0717xray} were the first to identify four major concentrations of large elliptical cluster galaxies, referred to as A, B, C, and D in their work. This scenario was later confirmed in a  parametric strong-lensing analysis by \citet{my0717}. More recently, \citet{diego0717}, exploiting the first third of the HFF data and using a non-parametric strong-lensing technique, has again found evidence for a quadri-modal mass distribution for the smooth component, i.e., once cluster members are removed. This agreement between the results from both parametric and non-parametric, grid-based modelling approaches (the latter having much more freedom than the former) is further evidence in favour of the four-component model. Like all the above-mentioned strong-lensing studies of \lens, ours also uses the original labelling by \citet{ma0717xray} of these four components (see Fig.~\ref{massmap}). In fact, we assume in this paper that the mass distribution of \lens\ is quadri-modal.

%Not only the number of mass components is of interest for structure formation and evolution scenario, but also
%the \emph{shape} of the different components matters. Indeed, if numerical simulations suggest an NFW-like mass profile
%for isolated unimodal structures, it is not expected that central cusps actually survive violent merging processes
%[check Rocha etal. and references within].
%%% DISCUSSION %%%%

All our results use the $\Lambda$CDM concordance cosmology with $\Omega_{\rm{M}} = 0.3, \Omega_\Lambda = 0.7$, and a Hubble constant \textsc{H}$_0 = 70$ km\,s$^{-1}$ Mpc$^{-1}$.  At the redshift of \lens\, this cosmology implies a scale of 6.4\,kpc/$\arcsec$. Magnitudes are quoted in the AB system. Figures are aligned with the conventional equatorial axes, i.e. north is up, and east is left.

\section{Multiple images}

\subsection{Previous work}

Prior to the HFF observations, 16 multiple-image systems, comprising 48 individual images, had been reported \citep{johnsonhff,richard14,coe15}. The analysis by \citet{diego0717}, using the first third of the data from the HFF observations, identified an additional 17 multiple-image systems, as well as 10 elongated features, assumed to be single images of lensed background galaxies.

More recently, at the same time as this paper, \citet{oguri0717} has presented a mass model
for \lens\, based on the full depth of the HFF observations.
We briefly discuss their findings in Section~\ref{comparRMS}.

\subsection{This work}

In this work, we revisit the strong-lensing identifications by \citet{diego0717}, agree with most of them (discarding, however, their system 30), and, using the full depth of the HFF observations, add 28 new multiple-image systems. The HFF observations thus enabled the discovery of 45 new systems (consisting of 117 images), bringing the grand total to 61 multiple-image systems in \lens, comprised of 165 individual images. Here we use as constraints only a subset of 132 multiple images that we consider to be the most reliably identified ones.

\begin{figure*}[ht!]
\begin{center}
\includegraphics[scale=0.53,angle=0.0]{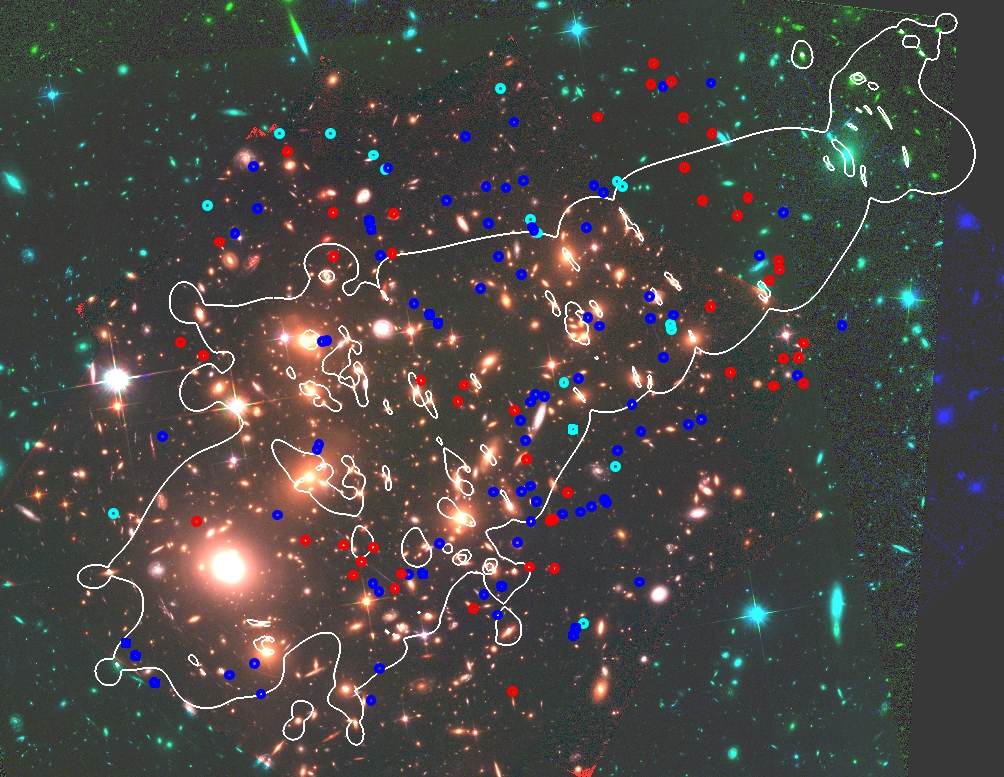}\\
\caption{Colour image (240$\arcsec$\,$\times$\,184$\arcsec$) of \lens\ based on HST images in the F435W, F606W, and F814W ACS pass bands. Multiple-image systems used in this work are marked: in red, the 48 multiple images identified prior to the HFF observations; in blue, the 84 images discovered thanks to the HFF observations,; in cyan, the 33 candidate images discovered in the HFF observations. Shown in white is the critical line for a source redshift of 7.}
\label{multiples}
\end{center}
\end{figure*}

For clarity, we present an overview of these images in two tables: Table~\ref{multipletable1} lists the multiple images known before the HFF observations and is thus identical to the one published in \citet{richard14}; the corresponding strong-lensing features are shown in red in Fig.~\ref{multiples}. Table~\ref{multipletable2} lists the multiple images discovered thanks to the HFF observations; they are shown in blue and cyan in Fig.~\ref{multiples}.  Where relevant, we use the notation of \citet{diego0717}.

We note that we have not been able to securely identify counter-images for some systems. We also report two radial arcs (systems 27 and 37). It is very likely that more multiple-image systems are present in the HFF data of \lens.

\subsection{Comparison with Diego et~al.}

\citet{diego0717} identified ten elongated features, interpreted as single-image arclets. These arclets are of potential interest in the region beyond the Einstein radius where strong-lensing constraints largely disappear. Combined with photometric estimates of their redshifts, the shape of these features 
(elongation, orientation) may provide valuable constraints. However, in our analysis, we do not use these features.

We agree with the identifications by \citet{diego0717}, except for the following few cases:

\begin{itemize}
\item image 25.3: we removed this image on the grounds of unclear morphology and colour. Adding 25.3 leads to a larger RMS for this system (2\arcsec instead of 1.3\arcsec). 
\item image 29.3: as for 25.3 we find the morphology and colour of 29.3 poorly determined; adding 29.3 leads to a larger RMS for this system (2.6\arcsec \,instead of 1.9\arcsec).
\item system 30: we disagree with this identification. Image 30.1 as proposed by \citet{diego0717} is a faint long arc located between images 1.2 and 1.3. In \citet{my0717}, we previously interpreted this feature as the merged tail of 1.2 and 1.3. This merged tail corresponds to the counter image of the tail associated with 1.1, for which we measured a spectroscopic of 2.963 \citep[labelled 1.1$^*$ in][]{my0717}. In addition, our parametric mass model is not able to reproduce system 30 as proposed by \citet{diego0717}.
\item system 34: we propose an alternative identification for image 34.1 which significantly improves the RMS of this system (1.6\arcsec instead of 3.7\arcsec when using the identification by \citet{diego0717}). We note that \citet{diego0717} has already reported a possible problem with this system. The identification we propose is supported by the geometry of systems 50 and 56.
\item system 39: \citet{diego0717} consider image 39.1 an elongated single-image feature. We associate this image with another one located on the other side of the arc, forming system 39. A third image is predicted, at least 1 magnitude fainter. Although we identify several candidates for this third image, none is sufficiently compelling, causing us to use only images 39.1 and 39.2 for this system.
%\item image 50.3: as for images 25.3 and 29.3; not sure of it; need to try again [CPU-CPU]
\end{itemize}

\subsection{Photometric redshifts}
In addition to the Frontier Fields observations, we include WFC3/UVIS (F275W, F336W) photometry in our analysis in order to estimate photometric redshifts for the multiple images. The corresponding data were obtained for $\it{HST}$ program ID 13389 (PI: B.Siana), which obtained deep UV imaging (8 orbits per filter) of three Frontier Field clusters (Abell 2744, \lens, and MACS J1149.5+2223). A complete description of the data reduction and photometry catalogues will be given in Siana et al.\ (in preparation). Here, we briefly summarize the photometric redshift measurements; a more detailed discussion will be presented in a forthcoming paper (Alavi et al., in preparation). 

We derived photometric redshifts of our galaxies using  the EAZY software package \citep{eazy} with the P\'{E}GASE \citep{pegase} stellar synthetic templates and a ${\chi^{2}}$ fitting procedure. When doing so, we included an additional spectral energy distribution (SED) template of a dusty starburst SED. EAZY parameterizes absorption from the intergalactic medium following the description presented in Madau (1995). We did not use the magnitude priors in EAZY, as our lensed galaxies are much fainter than the luminosity range covered by the priors.

We note that the resulting photometric redshift estimates (listed in Tables~\ref{multipletable1} and \ref{multipletable2}) are not used as constraints in the strong-lensing analysis; they are, however, very helpful for associating images with each other while looking for multiple-image systems.

\section{Strong-lensing analysis}

\subsection{Arc spectroscopy}

Previous work \citep{my0717,glass0717,vanzella,glass} reported spectroscopic redshifts for nine multiple-image systems (systems 1, 3, 4, 6, 12, 13, 14, 15 and 19). In addition, a spectroscopic redshift of $z{=}5.51$ was measured for system 68 (Clement et~al., in preparation), bringing to ten the total number of multiple-image systems in \lens\ for which spectroscopic redshifts are available. 

\citet{glass} reported a "probable" redshift of $z{=}0.928$ for image 5.1. We disregard this value.
Given the colour and morphology of the three images constituting system
5, we are confident in this multiply imaged system, whose geometrical configuration does not support a low redshift as reported by \cite{glass}. 
Indeed, our model predicts a redshift around 4 (Table~\ref{multipletable1}), in agreement with the 
predictions from \citet{diego0717} and \citet{oguri0717}. 
In addition, a non-detection in all filters bluer than the B-band can suggest a high-redshift
solution ($z>3$), as this possibility is also favoured by the CLASH survey which estimates
a photometric redshift of 4.5 for images 5.1 and 5.2 \citep{jouvel2014}.
%%%%%%% ZZZZ a few words by anahita ? %%%%%%%%%%%%%%

\subsection{Methodology}
As in our previous work \citep[see, \emph{e.g.}][]{mypaperIII}, our mass model consists of large-scale dark matter (DM) haloes, whose individual mass is larger than that of a galaxy group (typically of the order of 10$^{14}$\,M$_{\sun}$ within 50$\arcsec$), and perturbations associated with individual cluster galaxies. 
For our analysis of \lens, we consider the 90 most luminous galaxies in the ACS field as perturbers
(corresponding to a magnitude limit equal to 23.4). We characterize these mass components (both on the cluster and the galaxy scale) with dual Pseudo Isothermal Elliptical Mass Distribution \citep[dPIE,][]{mypaperI,ardis2218}, parametrized by a fiducial velocity dispersion $\sigma$, a core radius $r_{\rm core}$, and a scale radius $r_{\rm s}$. For the individual cluster galaxies, empirical scaling relations (without any scatter) are used to derive their dynamical dPIE parameters (central velocity dispersion and scale radius) from their luminosity (the core radius being set to a vanishingly small value of 0.3 kpc), whereas all geometrical parameters (centre, ellipticity, position angle) are set to the values measured from the light distribution. 
More precisely, the scaling relations are given by:
\begin{equation}
r_{\mathrm{s}}=r^*_{\mathrm{s}} (\frac{L}{L^*})^{\frac{1}{2}} \quad \& \quad
\sigma=\sigma* (\frac{L}{L^*})^{\frac{1}{4}}  
\end{equation}
We allow the velocity dispersion of an $L^*$ galaxy to vary between 100 and 250 km\,s$^{-1}$, 
whereas its scale radius is forced to be less than 70 kpc, thus accounting for tidal stripping of galaxy-size dark matter haloes \citep[see, \emph{e.g.}][and references therein]{mypaperII,mypaperIV,priya4,wetzel}. Model optimisation is performed in the image plane using the \textsc{lenstool}\footnote{http://www.oamp.fr/cosmology/lenstool/} software \citep{jullo07}.

\paragraph{Positional uncertainty.}
The positional uncertainties of the images is an important ingredient for the
$\chi^2$ computation. They affect the derivation of errors,
in the sense that smaller positional uncertainties will, in general, result in smaller statistical uncertainties, which may in fact be underestimated.

In principle, deep HST images like the ones used in this paper allow compact images to be located to an astrometric precision of the order of 0".05. However, parametric cluster lens models often fail to reproduce or predict image positions to this precision, yielding instead typical image-plane RMS values between 0.2" and a few arc seconds. Nominal positional uncertainties of 1.4" are usually chosen \citep{adiclash} to account for the contribution of fore- or background structures that are not included in our simple mass models \citep{jullo2010,daloisio,host}. In this work, we use an even larger positional uncertainty of 2.0" (i.e., a value of the order of the image plane RMS), in order to attain a reduced $\chi^2$ of order 1.

Finally, we should keep in mind that we are trying to reproduce observed and, to some extent, poorly understood
structures located far away using rather simple parameterized mass models.

\subsection{A multimodal mass model}
Acknowledging previous results, we adopt and optimize a quadri-modal mass distribution for \lens. Each mass component is associated with one of the four components (A, B, C and D), its location set to coincide with the corresponding light peak (red diamonds in Fig.~\ref{massmap}). As in prior studies, we allow the positions of these components to vary within $\pm$20\arcsec of the associated light peak, while the ellipticity is limited to a range from 0 to 0.7 (in units of $(a^2-b^2)/(a^2+b^2)$). The velocity dispersion of each component is allowed to vary between 400 and 1\,500\,km\,s$^{-1}$, and the core radius may take any value between 1 and 30\arcsec. Finally, the scale radius --- unconstrained by our data --- is fixed at 1\,000\arcsec.
%Results are given in the next Section.

\section{Mass distribution from strong lensing}

\subsection{Cored mass components: A flat DM distribution?} 
Our quadri-modal mass model is able to reproduce the lensing constraints with an image-plane RMS of 1.9\arcsec. The resulting parameters of all mass components are given in Table~\ref{tableres}; the corresponding mass contours are shown in yellow in Fig.~\ref{massmap}.

We find that the {\emph{total}} mass distribution (i.e., smooth DM component + galaxy-scale perturbers) follows the light distribution.
The centre of each component is offset from the associated light concentration by 19\arcsec, 27\arcsec, 13\arcsec, and 14\arcsec for components A, B, C, and D, respectively. In light of the large core radii of most of the four components (21\arcsec, 28\arcsec, 21\arcsec, and 5\arcsec for A, B, C, and D, respectively); however, these offsets are not necessarily significant. Specifically, our analysis does not support the hypothesis advanced by \citet{diego0717} that component A is significantly offset from the closest light peak.
These authors argued that this offset is probably due to the lack of lensing constraints in that area.

Since three of the four components have core radii larger than 15\arcsec, i.e., larger than 100\,kpc at the redshift of \lens, their superposition leads to a relatively flat mass distribution. In the following, this model is referred to as the \emph{cored} mass model.

Various mechanisms that could lead to such large cores have been proposed \citep[see also][for a thorough discussion]{diego0717,diego2015c}. Large cores in massive central galaxies are thought to be caused by massive black holes that clear the centre of galaxies of stars \citep[see, \emph{e.g.}][]{lopezcruz}. This may affect dark matter as well, but the scales on which this mechanism can be expected to be effective (less than 10 kpc) are smaller than the scale we are interested in here ($\sim$ 100 kpc). Actif galactic nucleus feedback can also contribute to the flattening of the central region \citep{martizzi12}, but once again this mechanism is efficient only on small scales (up to 10--15 kpc). On scales of 100 kpc, large cores may result from violent interactions related to merging events in this actively evolving cluster, although several studies find that DM profiles are not strongly altered by a collision \citep{ricker_sarazin,dekel,molnar}. Finally, heating of massive galaxies by dynamical friction against the diffuse dark matter distribution of the cluster can flatten the slope of the DM density profile, and sometimes even dominate over adiabatic contraction \citep[see, \emph{e.g.}][]{elzant01,elzant04}. Another explanation might be self-interacting dark matter particles \citep{priya08,rocha2013}, known to flatten the cusps of cluster scale haloes on scales up to 100 kpc, or the multi-coupled Dark Energy scenario \citep{garaldi}.

Given the implications of a flat mass profile for \lens, we investigate its mass distribution further by testing whether large-core components are really required by the data. To this end, we explore in the next subsection a peaky, non-cored mass model.

\begin{table*}
\begin{center}
\begin{tabular}[h!]{cccccccc}
\hline
\hline
\noalign{\smallskip}
Component  & $\Delta$ \textsc{ra}  & $\Delta$ \textsc{dec} & $e$ & $\theta$ & r$_{\mathrm{core}}$ (\arcsec) &  r$_{\mathrm{s}}$ (\arcsec)   &$\sigma$ (\footnotesize{km\,s$^{-1}$})\\ 
\noalign{\smallskip}
\hline
\noalign{\smallskip}
\noalign{\smallskip}
C &  -9.1$^{+1.5}_{-0.5}$  & -8.7 $\pm$ 1.4  &  0.34 $\pm$ 0.03 &  54 $\pm$ 4 &  16.1 $\pm$ 2.0 & [1000] & 895 $\pm$ 29.0 \\
\noalign{\smallskip}
\hline
\noalign{\smallskip}
D &  26.6 $\pm$ 0.9  & -19.3 $\pm$ 1.1 & 0.65 $\pm$ 0.03 &  52 $\pm$ 3  &  4.5 $\pm$ 0.8  & [1000] & 494$\pm$15  \\
\noalign{\smallskip}
\hline
\noalign{\smallskip}
B &  37.3 $\pm$ 3.9  & 34.9 $\pm$ 1.8  &  0.50 $\pm$ 0.06  &  3$^{+9}_{-2}$  &  27.6 $\pm$ 2.1 & [1000] & 800 $\pm$ 38 \\
\noalign{\smallskip}
\hline
\noalign{\smallskip}
A &  114.9 $\pm$ 2.9  & 64.7 $\pm$ 1.1  & 0.68 $\pm$ 0.01 &  9 $\pm$ 5  &  20.7 $\pm$ 2.1  & [1000]  & 880 $\pm$ 22  \\
\noalign{\smallskip}
\hline
\noalign{\smallskip}
L$^*$ galaxy & --  & --  & --    &    --   &  [0.05] &  8.3 $\pm$ 0.7  & 250.3 $\pm$ 8.4 \\
\noalign{\smallskip}
\hline
\hline
\end{tabular}
\caption{dPIE parameters inferred for the dark matter components considered in the optimization procedure: the four large-scale dark matter clumps (A, B, C, D)
and the galaxy-scale component. These parameters correspond to the \emph{cored} mass model.
Coordinates are given in arcseconds with respect to
$\alpha=109.3982, \delta=37.745778$.
The ellipticity $e$ is for the mass distribution.
Error bars correspond to $1\sigma$ confidence level.
Parameters in brackets are not optimized.
For the scaling relations, the zero point is set
to a magnitude equal to 20.66.
}
\label{tableres}
\end{center}
\end{table*}

\begin{table*}
\begin{center}
\begin{tabular}[h!]{cccccccc}
\hline
\hline
\noalign{\smallskip}
Component  & $\Delta$ \textsc{ra}  & $\Delta$ \textsc{dec} & $e$ & $\theta$ & r$_{\mathrm{core}}$ (\arcsec) &  r$_{\mathrm{s}}$ (\arcsec)   &$\sigma$ (\footnotesize{km\,s$^{-1}$})\\ 
\noalign{\smallskip}
\hline
\noalign{\smallskip}
\noalign{\smallskip}
C &  -0.9 $\pm$ 0.5  & 4.9$^{+0.1}_{-0.5}$  &  $>$ 0.38 &  63 $\pm$ 2 & 4.9$^{+0.1}_{-0.2}$ & [1000] & 889 $\pm$ 11  \\
\noalign{\smallskip}
\hline
\noalign{\smallskip}
D &  31.8 $\pm$ 0.5 & -14.5 $\pm$ 0.5 & $>$ 0.69 & 46 $\pm$ 3 &  4.8 $\pm$ 0.2  & [1000] & 617 $\pm$ 8  \\
\noalign{\smallskip}
\hline
\noalign{\smallskip}
B &  64.9 $\pm$ 1.0  & 39.9$^{+0.1}_{-0.2}$  &  $>$ 0.54  &  173 $\pm$ 18 &  4.9$^{+0.1}_{-0.6}$ & [1000] & 733 $\pm$ 13 \\
\noalign{\smallskip}
\hline
\noalign{\smallskip}
A &  134.9$^{+0.1}_{-0.7}$  & 75.0$^{+0.5}_{-0.1}$  & 0.61 $\pm$ 0.03 &  12.0 $\pm$ 9  &  2.7 $\pm$ 0.4  & [1000]  & 795 $\pm$ 16  \\
\noalign{\smallskip}
\hline
\noalign{\smallskip}
L$^*$ galaxy & --  & --  & --    &    --   &  [0.05] &  11.4 $\pm$ 1.4  & 161 $\pm$ 6\\
\noalign{\smallskip}
\hline
\hline
\end{tabular}
\caption{Same as Table~\ref{tableres} for the \emph{non-cored} mass model.
}
\label{tableres2}
\end{center}
\end{table*}

\begin{table*}
\begin{center}
\begin{tabular}[h!]{cccccccc}
\hline
\hline
\noalign{\smallskip}
Component  & $\Delta$ \textsc{ra}  & $\Delta$ \textsc{dec} & $e$ & $\theta$ & c$_{\mathrm{200}}$ &  r$_{\mathrm{s}}$ (\arcsec)   &$\sigma$ (\footnotesize{km\,s$^{-1}$})\\ 
\noalign{\smallskip}
\hline
\noalign{\smallskip}
\noalign{\smallskip}
C &  -2.3 $\pm$ 0.6 & -0.7 $\pm$ 0.8  &  0.13 $\pm$ 0.02 &  53 $\pm$ 4 &  3.4 $\pm$ 0.2 & 81.8 $\pm$ 6.0 & -- \\
\noalign{\smallskip}
\hline
\noalign{\smallskip}
D &  32.6 $\pm$ 0.4  & -13.5 $\pm$ 0.5 & 0.55 $\pm$ 0.03 &  50 $\pm$ 2  &  5.3 $\pm$ 0.3  & 25.3 $\pm$ 2.2 & --  \\
\noalign{\smallskip}
\hline
\noalign{\smallskip}
B &  63.3 $\pm$ 1.2  & 44.6 $\pm$ 0.4  &  0.51 $\pm$ 0.06  &  5 $\pm$ 2  & 3.0$^{+0.1}_{-0.0}$ & 63.9 $\pm$ 3.7 & -- \\
\noalign{\smallskip}
\hline
\noalign{\smallskip}
A &  133.7 $\pm$ 0.9  & 77.4 $\pm$ 1.4  & 0.23 $\pm$ 0.04 &  165 $\pm$ 15  &  5.5 $\pm$ 0.3  & 32.5 $\pm$ 0.35  & --  \\
\noalign{\smallskip}
\hline
\noalign{\smallskip}
L$^*$ galaxy & --  & --  & --    &    --   &  -- &  9.2 $\pm$ 0.4  & 188 $\pm$ 9 \\
\noalign{\smallskip}
\hline
\hline
\end{tabular}
\caption{NFW parameters inferred for the four large-scale dark matter clumps (A, B, C, D) and the dPIE parameters for
the galaxy-scale component (core radius fixed to 0.3\,kpc). These parameters correspond to the NFW mass model.
Coordinates are given in arcseconds with respect to
$\alpha=109.3982, \delta=37.745778$.
The ellipticity $e$ is for the mass distribution.
Error bars correspond to $1\sigma$ confidence level.
Concerning the scaling relations, the zero point is set
to a magnitude equal to 20.66.
}
\label{tableres3}
\end{center}
\end{table*}

\subsection{Non-cored mass components: A peaky DM distribution?}

The differences between the \emph{non-cored} and the \emph{cored} mass models lie in the {\emph{shape and location}} of each mass component:  the position of each mass component is required to lie within 5\arcsec of the associated light peak in both right ascension and declination, and its core radius is forced to be smaller than 5\arcsec. Interestingly, we find that this \emph{non-cored} mass model is also able to reproduce the observational constraints, with an image-plane RMS of 2.4\arcsec, similar to that of the \emph{cored} mass model. 
Therefore, both models fit the data equally well, in the sense that they give
comparable total RMS.

Best-fit parameters for this mass model are listed in Table~\ref{tableres2}.
We can appreciate that they differ significantly from the best-fit parameters obtained for the
\emph{cored} mass model.
The corresponding mass contours are shown in green in Fig.~\ref{massmap}.
Essentially, the \emph{total mass} distribution of this model also follows the light distribution.

\subsection{NFW profile}

We further investigate the impact of the profile of the DM components on our model's ability to satisfy the lensing constraints by adopting an NFW profile \citep{nfw96}, parametrized by a scale radius $r_s$ and a concentration parameter c$_{200}$, for each of the four cluster-scale components.
We note that this exercise is not physically motivated. Although NFW profiles are well suited to describe isolated, relaxed DM haloes, they are not necessarily expected to be adequate for the parameterization of this complex merging system. While the four subclusters may have had an NFW mass profile before their collision, it is not obvious that they will keep this shape during the violent merging process acting in \lens.

Here we test two NFW models. In the first, the location of each mass component is allowed to vary within $\pm$20$\arcsec$ of its associated light peak (the same limits used for the \emph{cored} mass model); in the second, this limitation is tightened to 5$\arcsec$ (the same limits used for the \emph{non-cored} mass model). In both cases, the scale radius is allowed to vary between 80 and 750 kpc, and the concentration parameter between 3 and 6.

We find that the best-fit parameters derived for either NFW model agree with each other. Even when the position of the components is allowed to deviate from that of the corresponding light peak, the best-fit position ends up consistent with the light distribution. We conclude that an NFW cusp without a luminous counterpart is not favoured by the data. In light of this result, we thus only discuss further the second NFW model. It too is able to reproduce the observational constraints, with an image-plane RMS of 2.2$\arcsec$, comparable to that of both the \emph{cored} and \emph{non-cored} models. Best-fit parameters for this mass model are listed in Table~\ref{tableres3}; the corresponding mass contours are shown in magenta in Fig.~\ref{massmap}.

This test thus further confirms that a peaky DM distribution for the large-scale mass components can accommodate the strong lensing
constraints.
%As mentioned before, we do not feel comfortable to describe the DM distribution in \lens\, using an NFW mass profile. Therefore, in the following, we will not consider the NFW model any more.

\subsection{A Hybrid Mass Model}
A key assumption of all models so far has been that each mass component is associated with one of the four light peaks in \lens\ (we note that this hypothesis is supported by previous studies, in particular by the series of ``blind tests" performed in \citet{my0717}). Consistent with this assumption, Fig.~\ref{massmap} shows the light distributions corresponding to the mass components A, C, and D to be dominated by a bright elliptical galaxy; however, this is not the case for mass component B, whose light peak coincides with a group of small elliptical galaxies. Acknowledging this peculiarity, we consider a hybrid mass model that combines some of the properties of the \emph{cored} and the \emph{non-cored} mass models. Components A, C, and D are described by a dPIE mass profile, whose right ascension and declination are required to remain within $\pm$5$\arcsec$ of the associated light peak and whose core radius is forced to be smaller than 5$\arcsec$, whereas component B is modelled by a dPIE mass profile whose position is allowed to vary within $\pm$20$\arcsec$ of the associated light peak, and whose core radius is allowed to reach 35$\arcsec$. Not surprisingly, this hybrid model is also able to reproduce the observational constraints with an RMS of 2.2$\arcsec$. The best-fit location of component B is 8.5$\arcsec$ away from the associated light peak, and its best-fit core radius is 20$\pm$2 $\arcsec$.

\section{Model comparison}

The small difference in RMS between the \emph{cored} and the \emph{non-cored} model (0.5\arcsec) suggests that the observational constraints, \emph{even in the HFF era}, cannot discriminate between a flat and a peaky dark matter distribution for the smooth component. This difference in RMS is comparable to or smaller than that due to an image mis-identification \citep[see, \emph{e.g.} the case of image 3.3 in Abell~2744 presented in][]{jauzac2744}, or to the difference caused by an unaccounted for structure along the line of sight \citep[\emph{e.g.}][]{host}, two effects that may also affect our analysis.

\subsection{Mass maps}
Fig.~\ref{compare_smooth} shows that the total mass maps generated by these two mass models are quite similar (top row) and follow the light distribution. In the middle and bottom row of Fig.~\ref{compare_smooth} this comparison is decomposed into the smooth component and the galaxy-scale component. In that decomposition, one can see important differences between the smooth components. It is peaky in the \emph{non-cored} mass model, and the four mass components are clearly visible. By contrast, the dark matter distribution of the \emph{cored} mass model is much more diffuse, leading to a very shallow mass profile. However, this effect is compensated for by the galaxy-scale component, which is much more massive in the \emph{cored} mass model (lower plots of Fig.~\ref{compare_smooth}).

\subsection{Convergence profile of the smooth component}
Fig.~\ref{comparmodel} (top left) compares the resulting convergence profile of the smooth component for  the two models. Following \citet{diego0717}, we take the centre of the profile to be the position of the most massive galaxy that is closest to the centre of group C ($\alpha=109.3982, \delta=37.745778$).  As expected from Fig.~\ref{compare_smooth}, the profiles differ visibly. For the \emph{cored}  mass model, the convergence profile is basically flat out to 20\,kpc and then decreases monotonically, whereas the profile for the \emph{non-cored} model features bumps corresponding to the mass components D, B, and A. For comparison, the same figure also shows the convergence profile for the smooth component inferred by \citet{diego0717}. It resembles the profile of our \emph{cored} model, but remains very flat to a much larger radius ($\sim$ 100\,kpc) before decreasing monotonically.

\subsection{Total mass profile}

\begin{figure*}[ht!]
\begin{center}
\includegraphics[scale=0.4,angle=0.0]{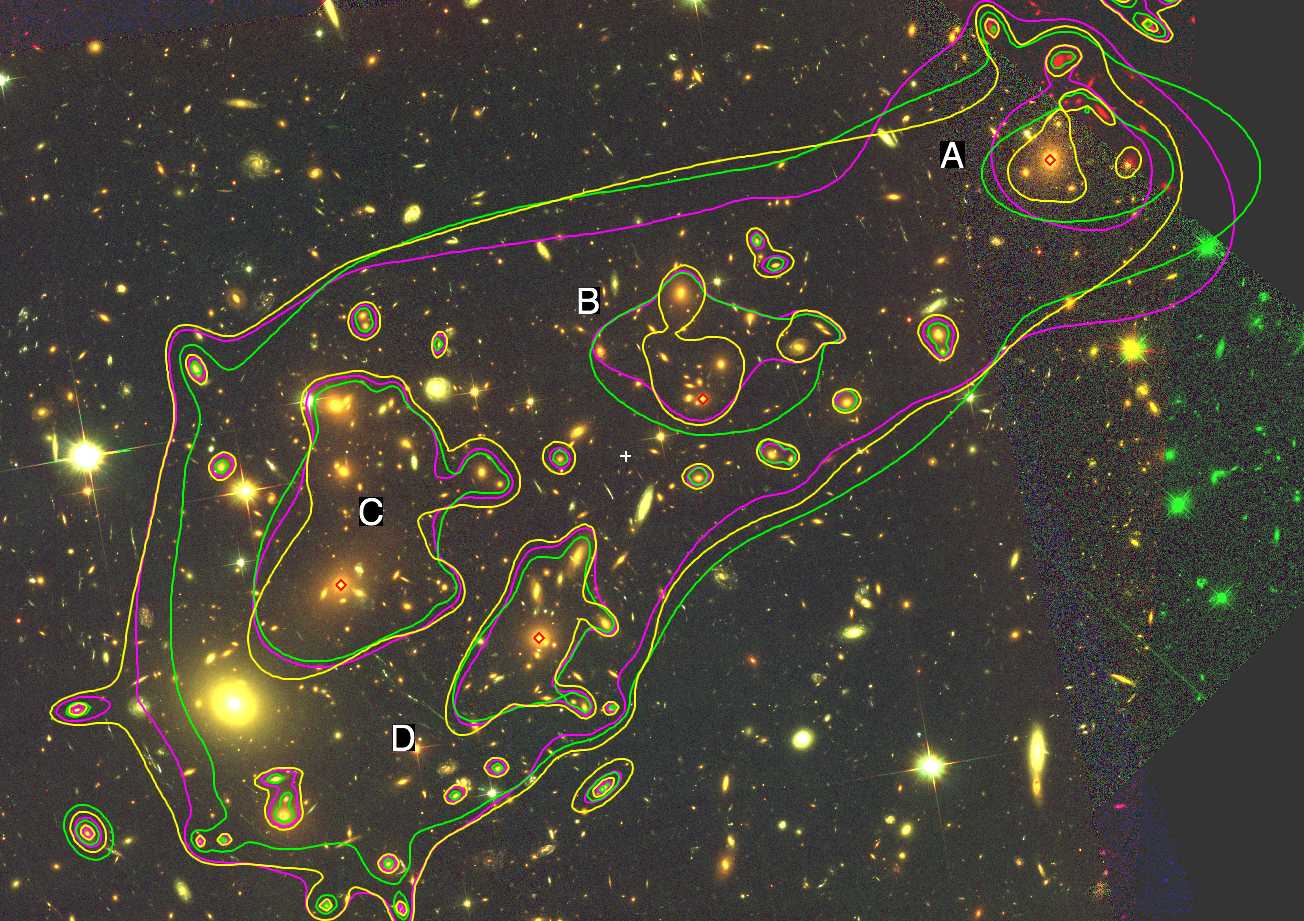}
\caption{Mass maps inferred from the \emph{cored} (yellow), the \emph{non-cored} (green) and the NFW (magenta) models. Contours show where the surface mass density equals 4, 7 $\times\,10^{10}$\,M$_{\sun}$ arcsec$^{-2}$.
%The location of each mass component is reported by an ellipse whose semi axes leanest
%corresponds to the errors on its location at the 1$\sigma$ confidence level: in magenta
%for the cored model and in cyan for the non-cored model.
Red diamonds correspond to the light peaks associated with each mass component. The white cross shows the barycentre of the Einstein ring as
estimated by \citet{massimo2011} at $\alpha=109.38002, \delta=37.752214$. The underlying image covers an area of 236\arcsec $\times$ 166\arcsec.
}
\label{massmap}
\end{center}
\end{figure*}

\begin{figure*}[h!]
\begin{center}
\vspace{1cm}
\includegraphics[scale=0.4,angle=0.0]{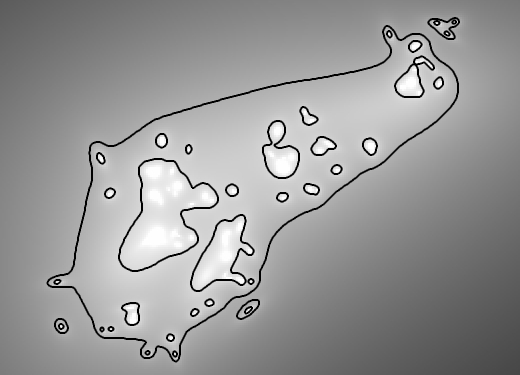}
\includegraphics[scale=0.4,angle=0.0]{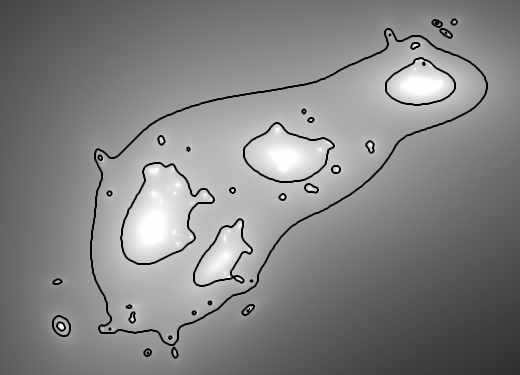}\\
\vspace{1cm}
\includegraphics[scale=0.4,angle=0.0]{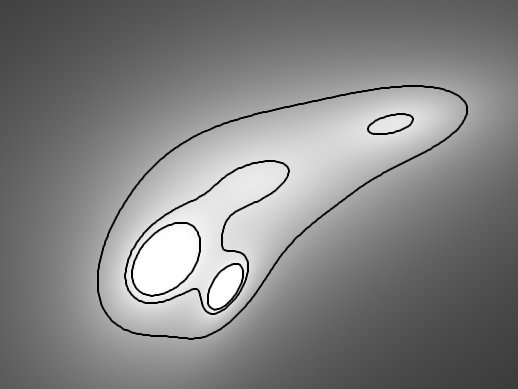}
\includegraphics[scale=0.4,angle=0.0]{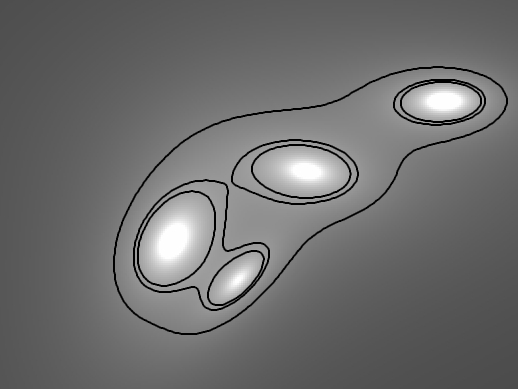}\\
\vspace{1cm}
\includegraphics[scale=0.4,angle=0.0]{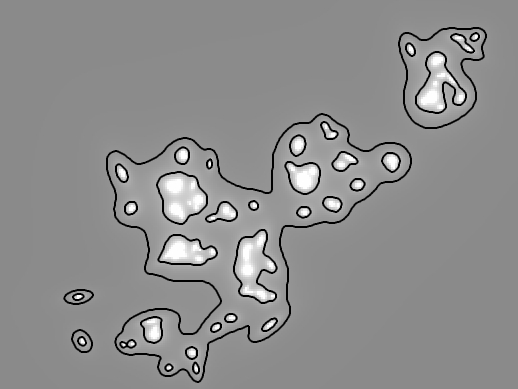}
\includegraphics[scale=0.4,angle=0.0]{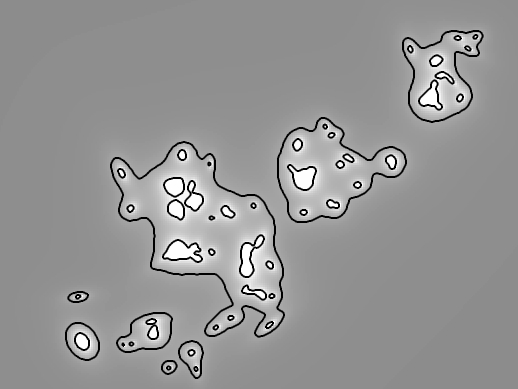}\\
\vspace{1cm}
\caption{
Comparison between the mass maps of the \emph{cored} (\emph{left}) and the \emph{non-cored} (\emph{right}) mass models. 
\emph{Top:} Total mass (smooth component + galaxies). Contours show where the surface mass density equals 4, 7 $\times\,10^{10}$\,M$_{\sun}$ arcsec$^{-2}$.
\emph{Middle:} contribution from the smooth component only. Contours delineate where the surface mass density equals 4, 5.5, 6.0 $\times\,10^{10}$\,M$_{\sun}$ arcsec$^{-2}$. \emph{Bottom:} Contribution from the galaxies component only. Contours mark where the surface mass density equals 0.5, 2 $\times\,10^{10}$\,M$_{\sun}$ arcsec$^{-2}$. We note that the galaxies gain more weight in the \emph{cored} mass model, in order to compensate
for the smoothness of the underlying large-scale mass components. Each panel measures 260\arcsec $\times$ 190\arcsec.
}
\label{compare_smooth}
\end{center}
\end{figure*}

\begin{figure*}[h!]
\begin{center}
\includegraphics[scale=0.39,angle=0.0]{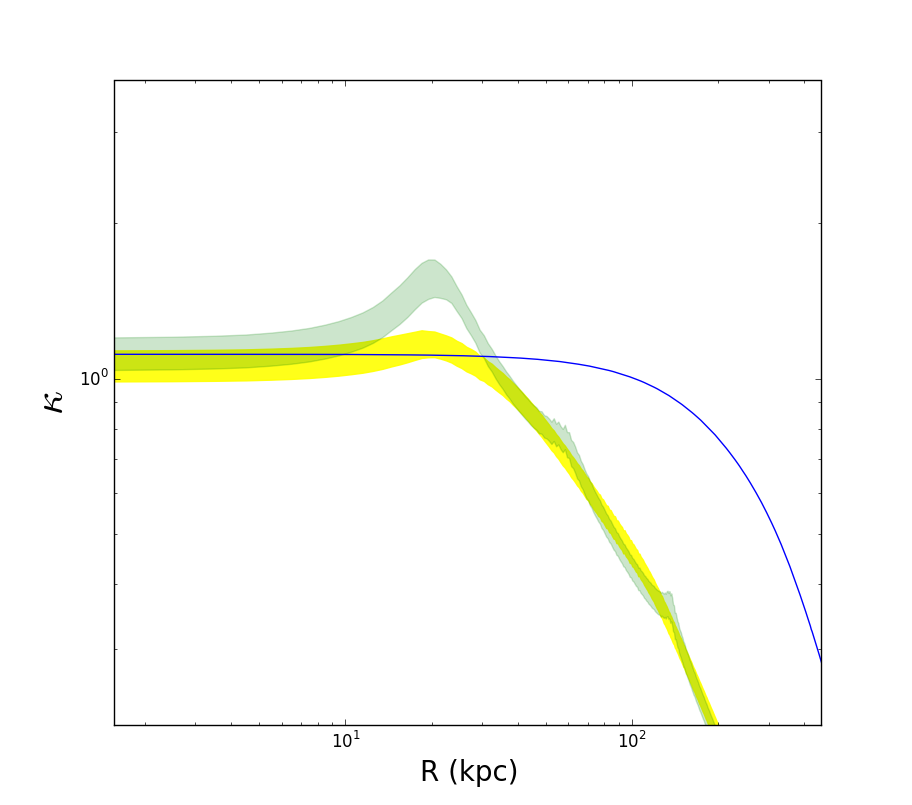}
\includegraphics[scale=0.39,angle=0.0]{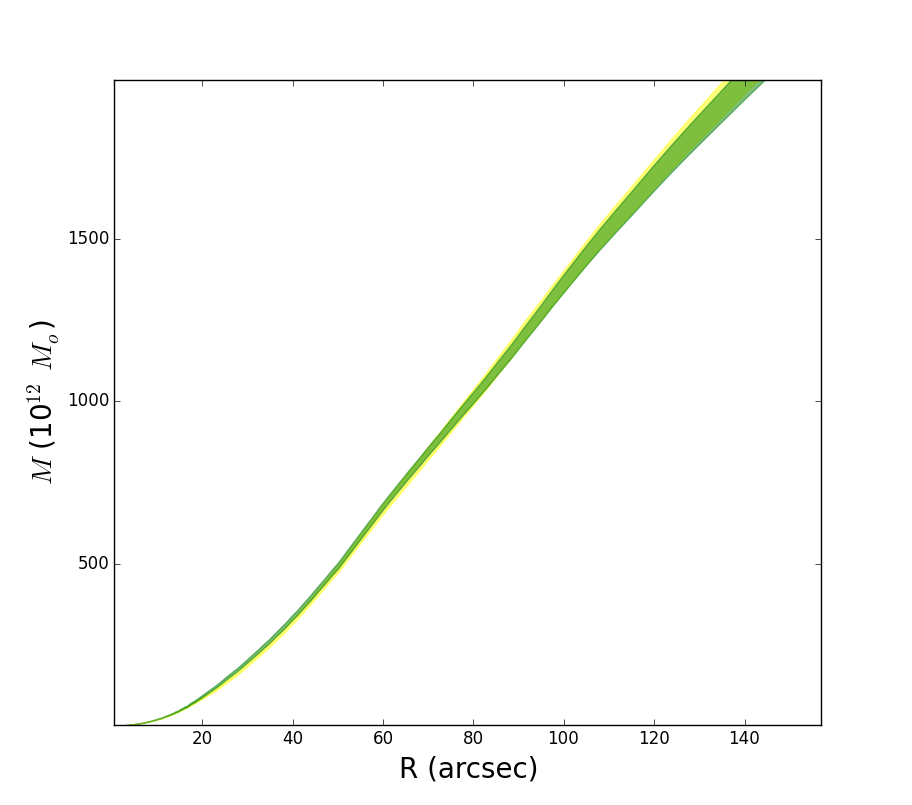}\\
\includegraphics[scale=0.4,angle=0.0]{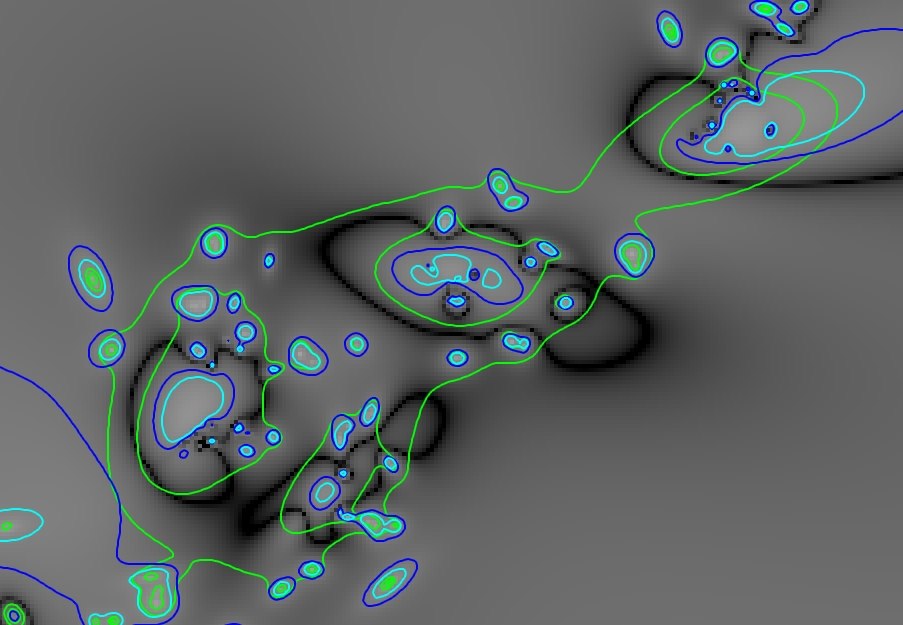}\\

\caption{\emph{Top left:} Surface-mass-density profile for the smooth component of the \emph{cored} (\emph{yellow}) and the \emph{non-cored} (\emph{green}) mass model. The blue line shows the same quantity, as derived by \citet{diego0717}). Shaded areas represent 3$\sigma$ uncertainties. \emph{Top right:} Cumulative two-dimensional mass profile for the \emph{cored} (\emph{yellow}) and the \emph{non-cored} (\emph{green}) mass model
(abscissa in arc seconds).
\emph{Bottom:} Absolute value of the relative difference between each models. Blue contours mark where the difference is 20\%, and cyan contours where it is 30\%. The mass contours corresponding to the \emph{non-cored} model are shown in green. The size of the field is 226$\arcsec$\,$\times$\,155$\arcsec$.
}
\label{comparmodel}
\end{center}
\end{figure*}

Although the large-scale (smooth) components of our two models results in noticeably different radial mass profiles, the models' cumulative total two-dimensional mass profiles are nearly indistinguishable (Fig.~\ref{comparmodel}, right).
Owing to the complex spatial distribution of the mass, the centre of \lens\ (needed to integrate the two-dimensional mass map) is not easily defined. Following \citet{massimo2011}, we choose the barycentre of the Einstein ring, at $\alpha=109.38002, \delta=37.752214$ (white cross on Fig.~\ref{massmap}), near the centre of the ACS frame. 
The projected mass within 990\,kpc (156$\arcsec$) of this location is found to be $M{=}(2.229 \pm 0.022)\times 10^{15}$ M$_{\sun}$ for the \emph{cored} model, and $M{=}(2.199 \pm 0.021)\times 10^{15}$ M$_{\sun}$ for the \emph{non-cored} model (1$\sigma$ statistical errors). As for other cluster mass models based on HFF data \citep{jauzac2744,0416hff,1149hff}, the errors thus imply 1\% precision. However, such claims disregard the additional, systematic uncertainty that we investigate here and which, based on the choice of mass model, leads to larger error bars on the mass. Averaging the two mass measurements yields $M{=}(2.214 \pm 0.037)\times 10^{15}$ M$_{\sun}$. It is likely that other systematic uncertainties, not taken into account here, further affect the reported value and its error.

A final comparison is presented in Fig.~\ref{comparmodel}, which shows the absolute value of the relative difference between the mass maps inferred for the \emph{cored} and the \emph{non-cored} model. We find differences of 30\% near massive cluster ellipticals 
and, importantly, also in the cores of the four large-scale components. 

\subsection{Comparison with other studies}
\label{comparRMS}
\citet{diego0717}, using the first third of the HFF observations, reported an image plane RMS of
2.8$\arcsec$. 
More recently, \citet{oguri0717}, using 173 images from the full depth of the HFF observations, 
reported an image plane RMS of 0.52$\arcsec$.
Their modelling approach is parametric. They place nine NFW components on the positions of bright
cluster members (of which six have a mass larger than 10$^{14}$\,$h^{-1}$\,M$_{\sun}$),
plus multipole perturbations modelling the external perturbations on the lens potential and the
asymmetry of the cluster mass distribution.
Compared to our approach, their model involves many more parameters, which may explain why their RMS
is significantly smaller than ours.
Their paper being focusses on the detection of high-redshift galaxies, 
and so modelling results 
that allow to pursue a more detailed comparison to be pursued are not provided.
In particular, it is not possible to see if their best mass model is closer to our
\emph{cored} or to our \emph{non-cored} mass model.

\section{Discussion}
Our analysis reveals that two very different models (in term of mass component parameters)
are able to reproduce all observational constraints equally well, in the sense that the total RMS is comparable. 
While we can state that the mass distribution in \lens\ is quadri-modal, further insights remain limited; in particular, we are not able to constrain the shape and the precise location of the four mass clumps, even with the exquisite data provided by the HFF project. Although this problem is likely exacerbated by the complexity of the mass distribution in \lens, it is likely to affect and limit our understanding of the mass distribution in other clusters too.

In the following we discuss ways to discriminate between these models.

\subsection{Stellar velocity dispersions}

As our two models lead to different mass distributions for the galaxy-scale component (see Fig.~\ref{compare_smooth}), here we 
investigate the role of the galaxy-scale perturbers in an attempt to find ways to discriminate between the two models.

The parameters characterizing the galaxy-scale perturbers, the velocity dispersion $\sigma$ and the scale radius $r_s$, differ noticeably: we find $\sigma{=}250$\,km\,s$^{-1}$ and $r_s{=}53$\,kpc for the \emph{cored} mass model, and $\sigma{=}161$\,km\,s$^{-1}$ and $r_s{=}73$\,kpc for the \emph{non-cored} mass model (for a magnitude of 20.66). Both of these parameters can be constrained independently by weak and strong galaxy-galaxy lensing in clusters, and by measurements of the stellar velocity dispersion from low-resolution spectroscopy. Recently, \citet{monna2015} conducted the latter type of investigation for the galaxies in Abell~383 and also presented a compilation of similar measurements in the literature (Fig.~11 of their paper). The parameter $\sigma$ is found to range from 80 to 300 km\,s$^{-1}$ and $r_s$ from 5 to 85 kpc, which are consistent with the values obtained for both of our two models.
Measurements of these parameters for a significant number of individual galaxies 
in \lens\, may allow us to discriminate between the \emph{cored} and the \emph{non-cored} mass model.
%Besides, it might be inaccurate to use priors for the galaxy scale component based
%on results obtained for other clusters.

\subsection{Redshift estimates and measurements}

In principle, the redshift predictions of either model might prove useful to discriminate between them. However,
all estimated redshifts predicted by the different mass models (listed in Tables~\ref{multipletable1} and \ref{multipletable2}) agree with each other at the 3$\sigma$ confidence level. We also note that for all systems but one (system 39), these model-based estimates agree with the photometric redshifts.

\subsection{Beyond strong lensing}

Using different probes beyond strong lensing might also allow us to at least alleviate the degeneracies between the cluster- and galaxy-scale mass components. Measuring the weak-lensing shear around cluster members is a very interesting avenue in this context, in particular in the flexion regime, which is well suited to probe galaxy-scale dark matter haloes.
%\citep{,,}.
Although flexion is challenging to measure, its signal has already been detected in much shallower
space-based images than the ones provided by the HFF \citep[see, \emph{e.g.}][on Abell~1689]{1689flexion}.

\section{Implications for magnification estimates}

The primary science goal of the HFF campaign is to use clusters as gravitational telescopes in order to probe deeper into the high-redshift Universe. Whether, or how well, this goal can be achieved depends critically on the accuracy with which one can recover the amplification that allows us to convert observed into intrinsic properties of said high-redshift objects.

Not surprisingly, the \emph{cored} and the \emph{non-cored} mass models presented in this paper lead to different magnification estimates, thus adding a systematic error that is in general larger than the statistical error derived from a single mass model. 
This systematic error is defined as the
difference between the amplification derived from the best-fit \emph{cored} model and
the amplification derived from the best-fit \emph{non-cored} model. It can sometimes be larger than the difference between the magnification values obtained by the different groups of modellers, using pre-HFF data. To quantify the effect for a given image, we compute the amplifications obtained by the different groups of modellers using pre-HFF data
\footnote{http://archive.stsci.edu/prepds/frontier/lensmodels/} and take the difference between the smallest and largest value as the ``modellers" uncertainty.

For a given source redshift, the implications of these systematic uncertainties for magnification estimates depends on \emph{where} one looks through the cluster. To illustrate this, for a few multiply imaged galaxies we list the following informations in Table~\ref{magtable}: the amplification with associated error bars and signal-to-noise ratio (in square brackets), for the \emph{cored} 
and the \emph{non-cored} models; the difference between these two 
amplifications (the systematic error developed in this paper), and the previously mentioned modellers uncertainty. The images used for this exercise were chosen to represent a wide range of magnifications: several are located far from the critical lines (images 19.3, 68.2, 71.2, and 61.2), one is close to the critical line in one model but not in the other (image 57.1), and another one lies close to the critical lines in both models (23.2).

In regions where the amplification is estimated with high confidence, we note how \emph{the amplification
difference between the \emph{cored} and the \emph{non-cored}} models is larger than the error associated with a given model and comparable
to the modellers uncertainty. In the case of image 57.1, we also note how the amplification is reasonably well constrained in the \emph{cored} model (signal-to-noise ratio of 2), but essentially unconstrained in the \emph{non-cored} model.

\begin{figure*}
\begin{center}
\includegraphics[scale=0.4,angle=0.0]{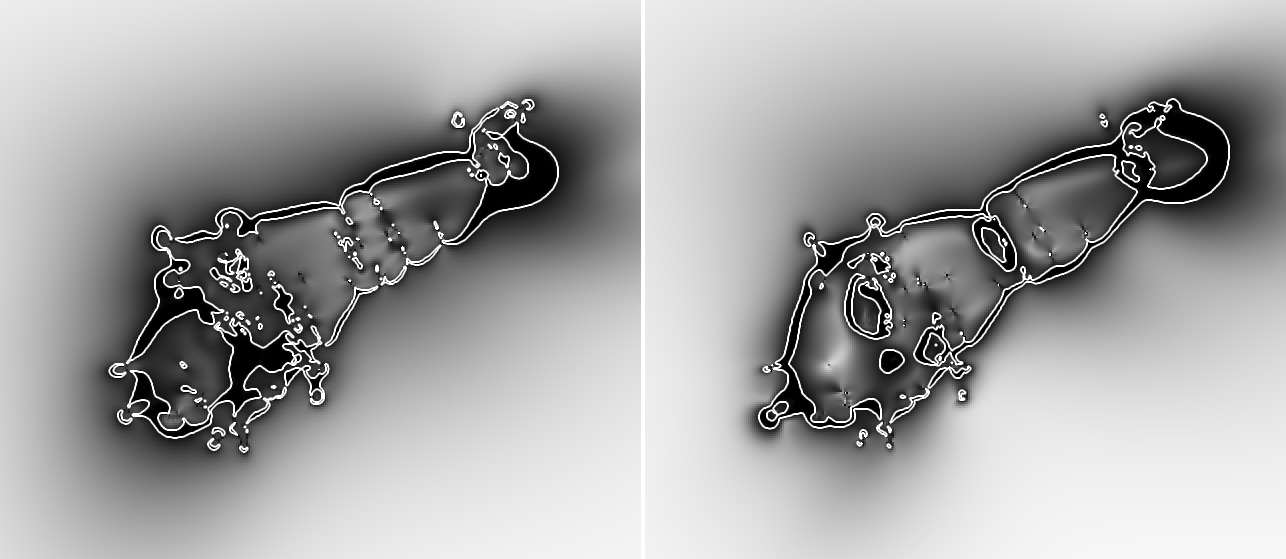}
\caption{Map (310$\arcsec$\,$\times$\, 310$\arcsec$) of the signal-to-noise ratio of the amplification, for the
\emph{cored} (\emph{left}) and the \emph{non-cored} (\emph{right}) mass models. White lines indicate where the signal-to-noise ratio equals 3. Darker areas interior to these lines correspond to regions where the signal-to-noise ratio is less than 3.} 
\label{snampli} 
\end{center}
\end{figure*}

This additional systematic uncertainty needs to be taken into account in all practical applications of a gravitational telescope, as it will decrease the area of the image plane where amplifications are well determined enough for credible studies of the high-redshift Universe. We quantify this effect by computing the area of the image plane for which the signal-to-noise ratio on the amplification is less than 3 for a source redshift of 7.  We find 1.1 and 1.3 arcmin$^2$ for the \emph{cored} and \emph{non-cored} models, respectively. When both models are considered valid descriptions, thus adding the above-mentioned systematic error, this area increases to 1.9 arcmin$^2$, i.e., by 50 and 70\%, respectively.
A graphic presentation of this change in area is provided by Fig.~\ref{snampli} which shows the signal-to-noise ratio for the \emph{cored} and the \emph{non-cored} mass models; the white line marks a value of 3.

\begin{table*}
\begin{center}
\begin{tabular}{ccccc}
\hline
ID & Ampli (cored) & Ampli (non-cored) & $\Delta$ [cored\,-\,noncored] & $\Delta$ [pre-HFF]  \\
\hline \\*[-1mm]
19.3 & 3.2 $\pm$0.15 [21]  & 2.8$\pm$0.07 [40] & 0.4  & 1.7  \\
68.2 & 24$\pm$2.0 [12] & 8.7$\pm$0.6 [14.5]  & 15.3  & 4.0  \\
71.2 & 12.0$\pm$1.4 [8.5] & 10.0$\pm$0.7 [14] & 2.0 & 63.5 \\ 
61.2 &  2.3$\pm$0.2 [11.5] & 1.6$\pm$0.1 [16] & 0.7 & 1.62 \\
57.1 & 12.3$\pm$6.0 [2] & 32.4$\pm$ 846 [$<$1] & $>100$ & 205 \\
23.2 & 1083$\pm$9167 [$<$1] & 21.8$\pm$2.5 [$<1$] & $>1000$ & 69 \\
\hline
\smallskip
\end{tabular}
\end{center}
\caption{
For different multiple images, we report: the value of the amplification with associated errors
and signal-to-noise ratio (in brackets), for the cored and the non-cored model; the difference in amplification between the cored and the
non-cored mass model; the difference in amplification considering pre-HFF data and the different
modelling/group software. Magnifications computed for $z=3.0$.}
\label{magtable}
\end{table*}

\section{Conclusion}

We present a strong-lensing analysis of \lens\ based on the full depth of the HFF data.

We find that the lensing constraints can be equally well satisfied by a mass model with a shallow large-scale DM component and one for which this component is peaky. Given the clear physical difference between the two models, we conclude that our \emph{insights into the DM distribution remain limited, even in the HFF era}. One way to discriminate between the two models would be to model the mass distribution while constraining simultaneously the large-scale smooth component and the galaxy-scale component, i.e. the two ingredients of our mass models that display generic degeneracies.

Our findings have important implications for magnification estimates. We show that the ambiguity between the two models leads to an additional systematic error that varies with position and needs to be taken into account when looking at the high-redshift Universe through \lens. Generally speaking, it decreases the area available for studying the high-redshift Universe by 50--70\%.

%\clearpage

\section*{Acknowledgement}

ML thanks Jose M.\ Diego for helpful exchanges during the writing of this paper. ML acknowledges the Centre National de la Recherche Scientifique (CNRS) for its support. This work was performed using facilities offered by CeSAM (Centre de donnéeS Astrophysique de Marseille-(http://lam.oamp.fr/cesam/). 
This work was granted access to the HPC resources of Aix-Marseille Universit\'e financed by the project Equip@Meso (ANR-10-EQPX-29-01) of the program "Investissements d'Avenir" supervised by the Agence Nationale pour la Recherche (ANR).
We acknowledge support from the Programme National de Cosmologie et Galaxie (PNCG). CG is grateful to CNES for financial support. This work was carried out with support of the OCEVU Labex (ANR-11-LABX-0060) and the A*MIDEX project (ANR-11-IDEX-0001-02) funded by the "Investissements d'Avenir" French government program managed by the ANR. MJ is supported by the Science and Technology Facilities Council [grant number ST/L00075X/1 \& ST/F001166/1]. This work used the DiRAC Data Centric system at Durham University, operated by the Institute for Computational Cosmology on behalf of the STFC DiRAC HPC Facility (www.dirac.ac.uk [www.dirac.ac.uk]), using equipment funded by BIS National E-infrastructure capital grant ST/K00042X/1, STFC capital grant ST/H008519/1, and STFC DiRAC Operations grant ST/K003267/1 and Durham University. DiRAC is part of the National E-Infrastructure. This study used gravitational-lensing models produced by PIs Bradac, Ebeling, Merten \& Zitrin, Sharon, and Williams funded as part of the HST Frontier Fields program conducted by STScI. STScI is operated by the Association of Universities for Research in Astronomy, Inc. under NASA contract NAS 5-26555. The lens models were obtained from the Mikulski Archive for Space Telescopes (MAST); we thank Dan Coe for kindly making them available to the community. 
PN acknowledges support from the National Science Foundation through grant AST-1044455.

\bibliographystyle{aa} % style aa.bst
%\bibliography{master}
\bibliography{draft}

\newpage
\appendix

\begin{table*}
\begin{center}
\begin{tabular}{cccccc}
\hline
ID & R.A. & Dec. & $z_{\rm spec}$ & $z_{\rm model}$ (\emph{non-cored}, \emph{cored}, NFW) & $z_{\rm phot}$\\
\hline \\*[-1mm]
%1.1$^{*}$ & 109.39587 & 37.74212 & 2.963 & -- \\ %strong Ly_alpha
1.1 & 109.39534 & 37.741178 & -- & --  & 3.07$\pm$0.06\\
1.2 & 109.39382 & 37.740095 & 2.963 & -- & 2.91$\pm$0.09\\
1.3 & 109.39098 & 37.738280 & 2.963 & -- &2.72$\pm$0.17\\
1.4 & 109.38436 & 37.736945 & --  & -- &3.11$\pm$0.06\\
1.5 & 109.40578 & 37.761378 & --  & -- &3.11$\pm$0.07\\
2.1 & 109.39281 & 37.741010 & --  & 2.6$\pm$0.3, 2.2$\pm$0.4, 3.0$\pm$0.3 &2.72$\pm$0.12\\ 
2.2 & 109.39043 & 37.739245 & --  & -- &2.72$\pm$0.07 \\ 
3.1 & 109.39855 & 37.741495 & 1.855  & -- & 1.87$\pm$0.04\\
3.2 & 109.39446 & 37.739176 & 1.855 & -- & 1.84$\pm$0.04\\
3.3 & 109.40715 & 37.753827 & 1.855  & -- & 1.84$\pm$0.04\\
4.1 & 109.38088 & 37.750127 & 1.855 & -- & 1.93$\pm$0.05\\
4.2 & 109.37644 & 37.744696 & 1.855  & --  & 1.84$\pm$0.04\\
4.3 & 109.39109 & 37.763296 & 1.855  & --  &1.82$\pm$0.04\\
5.1 & 109.37991 & 37.746861 & --  & 4.1$\pm$0.2, 3.8$\pm$0.2, 3.7$\pm$0.1  & --\\
5.2 & 109.37791 & 37.742810 & --  & -- &--\\
5.3 & 109.40003 & 37.767399 & --  & -- &--\\
6.1 & 109.36436 & 37.757091 & 2.393  & -- &2.37$\pm$0.06\\
6.2 & 109.36271 & 37.752693 & 2.393  & -- &2.27$\pm$0.07\\
6.3 & 109.37388 & 37.769711 & 2.393  & -- & --\\
7.1 & 109.36657 & 37.766343 & --  & 1.9$\pm$0.1, 2.0$\pm$0.1, 2.0$\pm$0.1 & -- \\
7.2 & 109.36505 & 37.764125 & --  & --& --\\
7.3 & 109.35905 & 37.751780 & --  & -- & 2.23$\pm$0.13\\  
8.1 & 109.36665 & 37.769694 & --  & 2.5$\pm$0.1, 2.9$\pm$0.2, 2.8$\pm$0.1 &--\\
8.2 & 109.36208 & 37.763125 & --  & -- &--\\
8.3 & 109.35652 & 37.751928 & --  & -- &--\\
12.1 &109.38516 & 37.751844 & 1.699  & -- &1.62$\pm$0.05 \\
12.2 &109.37762 & 37.742878 & 1.699  & -- &1.73$\pm$0.06\\
12.3 &109.39122 & 37.760626 & 1.699  & -- &1.71$\pm$0.05\\
13.1 &109.38567 & 37.750733 &2.547& -- & 2.47$\pm$0.06\\ 
13.2 &109.37756 & 37.739627 & --  & -- &2.54$\pm$0.07\\
13.3 &109.39621 & 37.763333 & 2.540  & -- &2.61$\pm$0.09\\
14.1 &109.38879 & 37.752163 & 1.855  & -- &1.84$\pm$0.04\\
14.2 &109.37966 & 37.739707 & 1.855  & -- &1.84$\pm$0.04\\
14.3 &109.39619 & 37.760427 & 1.855  & --  &1.84$\pm$0.02\\
15.1 &109.36766 & 37.772059 & 2.405 & -- & --\\
15.2 &109.35862 & 37.760127 & --  & --  &--\\
15.3 &109.35654 & 37.754641 & --  & --  & 3.40$\pm$0.08\\
16.1 &109.36916 & 37.773279 & --  & 3.2$\pm$0.3, 3.6$\pm$0.3, 3.1$\pm$0.1 &-- \\
16.2 &109.35856 & 37.759558 & --  & --  &--\\
16.3 &109.35694 & 37.753691 & --  & --  &--\\
17.1 &109.36938 & 37.771869 & --  & 2.7$\pm$0.2, 2.8$\pm$0.2, 2.7$\pm$0.1 &-- \\
17.2 &109.35938 & 37.758792 & --  & --  &--\\
17.3 &109.35822 & 37.753609 & --  & --  & 3.15$\pm$0.14\\
18.1 &109.36425 & 37.768628 & --  & 2.0$\pm$0.1, 2.6$\pm$0.5, 2.2$\pm$0.1 &-- \\
18.2 &109.36121 & 37.764326 & --  & --  &-- \\
19.1 &109.40906 & 37.754681 & 6.40 & -- &-- \\
19.2 &109.40772 & 37.742731 & 6.40 & -- &-- \\
19.3 &109.38105 & 37.731391 & 6.40 & -- &-- \\
\hline
\smallskip
\end{tabular}
\end{center}
\caption{Multiple-image systems found before the FF observations.
Uncertainties quoted for redshifts predicted by our model, $z_{\rm model}$,  correspond to the 1$\sigma$ 
confidence level. They are given for the \emph{non-cored}, the \emph{cored}, and the NFW models.
In the last column, we show the estimate of the photometric redshift, when possible, together
with the 1$\sigma$ error bars.}
\label{multipletable1}
\end{table*}

\begin{table*}
\begin{center}
\begin{tabular}{cccccc}
\hline
ID & R.A. & Dec. & $z_{\rm spec}$ & $z_{\rm model}$ (\emph{non-cored}, \emph{cored}, NFW) & $z_{\rm phot}$\\
\hline \\*[-1mm]
20.1 & 109.37420 & 37.7651450 & -- & 3.4$\pm$1.0, $>$2.6, uncons. &-- \\
20.2 & 109.37340 & 37.7646610 & -- & -- &-- \\
21.1* & 109.37241 & 37.746392 & -- & -- &2.61$\pm$0.16\\
21.2* & 109.37673 & 37.752029 & -- & -- & 2.65$\pm$0.12\\
22.1* & 109.36773 & 37.755897 & -- & -- & 0.87$\pm$0.11\\
22.2* & 109.36769 & 37.755556 & -- & -- & 0.93$\pm$0.05\\
23.1* & 109.37958 & 37.762879 & -- & -- & --\\
23.2* & 109.37897 & 37.761983 & -- & -- &-- \\
24.1 & 109.392290 & 37.732946 & -- & 2.7$\pm$0.1, 2.5$\pm$0.1, 2.5$\pm$0.1 &--\\
24.2 & 109.410560 & 37.748427 & -- & -- &--\\
25.1 & 109.380290 & 37.744746 & -- & 4.0$\pm$0.1, 3.8$\pm$0.1, 3.8$\pm$0.2 &--\\
25.2 & 109.379510 & 37.742762 & -- & -- &--\\
25.3 & 109.402910 & 37.766411 & -- & -- &--\\
27.1 & 109.397390 & 37.747883 & -- & 1.6$\pm$0.9, 3.0$\pm$1.1, uncons. & 2.95$\pm$0.18\\
27.2 & 109.397560 & 37.747567 & -- & -- & 2.17$\pm$0.32\\
29.1 & 109.400870 & 37.743174 & -- & 1.7$\pm$0.1, 1.7$\pm$0.1, 1.7$\pm$0.1 & 1.45$\pm$0.07\\
29.2 & 109.392840 & 37.738612 & -- & -- &1.62$\pm$0.06\\
31.1 & 109.374720 & 37.756347 & -- & 1.6$\pm$0.1, 1.6$\pm$0.1, 1.6$\pm$0.1 &1.55$\pm$0.08\\
31.2 & 109.371000 & 37.750538 & -- & -- &1.52$\pm$0.06\\
31.3 & 109.381610 & 37.764973 & -- & -- &1.57$\pm$0.08\\
32.1 & 109.369500 & 37.757729 & -- & 2.3$\pm$0.1, 2.8$\pm$0.1, 2.5$\pm$0.1 & --\\
32.2 & 109.380950 & 37.769382 & -- & -- & --\\
32.3 & 109.366250 & 37.749210 & -- & -- & 2.40$\pm$0.05\\
33.1 & 109.383770 & 37.758259 & -- & 3.5$\pm$0.3, 3.7$\pm$0.2, 3.7$\pm$0.1 &--\\
33.2 & 109.386620 & 37.764144 & -- & -- &--\\
33.3 & 109.370370 & 37.738712 & -- & -- &--\\
34.1 & 109.391580 & 37.766308 & -- & 2.6$\pm$0.2, 2.7$\pm$0.2, 2.6$\pm$0.1 &--\\
34.2 & 109.379110 & 37.751227 & -- & -- &1.03$\pm$0.05\\
34.3 & 109.373300 & 37.744211 & -- & -- &1.11$\pm$0.09\\
36.1 & 109.364330 & 37.771976 & -- &2.4$\pm$0.2, 2.9$\pm$0.2, 2.6$\pm$0.1 &--  \\
36.2 & 109.358240 & 37.763325 & -- & -- &--\\
36.3 & 109.353290 & 37.755808 & -- & -- &--\\
37.1 & 109.397090 & 37.754736 & -- & 3.8$\pm$0.8, 3.3$\pm$0.7, 3.3$\pm$0.4 &--\\
37.2 & 109.396720 & 37.754793 & -- & -- &--\\
39.1 & 109.402270 & 37.731234 & -- & 4.3$\pm$0.3, 4.1$\pm$0.3, 4.1$\pm$0.3 &1.82$\pm$0.04\\
39.2 & 109.404950 & 37.732519 & -- & -- & 2.14$\pm$0.19\\
39.3* & 109.414740 & 37.743276 & -- & -- &--\\
45.1 & 109.389820 & 37.739214 & -- & 2.7$\pm$0.1, 2.7$\pm$0.1, 2.6$\pm$0.1 &--\\
45.2 & 109.383480 & 37.737879 & -- & -- &--\\
45.3 & 109.404470 & 37.761960 & -- & -- &-- \\
49.1 & 109.402800 & 37.733260 & -- & $>$4.8, 4.1$\pm$0.3, 3.8$\pm$0.2 & 3.40$\pm$0.08\\
49.2 & 109.393000 & 37.730812 & -- & -- & 3.54$\pm$0.06\\
50.1 & 109.374440 & 37.743736 & -- & 3.7$\pm$0.4, 3.6$\pm$0.3, 3.6$\pm$0.2 & 3.19$\pm$0.10\\
50.2 & 109.379580 & 37.750708 & -- & -- &--\\
50.3* & 109.392800 & 37.767181 & -- & -- &--\\
52.1 & 109.368380 & 37.771761 & -- & 3.0$\pm$0.2, 3.3$\pm$0.2, 3.3$\pm$0.3 &--\\
52.2 & 109.360230 & 37.760497 & -- & -- &--\\
52.3 & 109.357040 & 37.752486 & -- & -- &--\\
55.1 & 109.373760 & 37.755778 & -- & 2.2$\pm$0.1, 2.4$\pm$0.1, 2.4$\pm$0.1 & --\\
55.2 & 109.370240 & 37.748743 & -- & -- & 2.17$\pm$0.11\\
55.3 & 109.385020 & 37.768411 & -- & -- & --\\
56.1 & 109.373150 & 37.744015 & -- & 3.2$\pm$0.2, 3.2$\pm$0.2, 3.0$\pm$0.1 &--\\
56.2 & 109.378400 & 37.751095 & -- & -- &--\\
56.3* & 109.391820 & 37.766216 & -- & -- &--\\
57.1 & 109.379030 & 37.744097 & -- & 1.7$\pm$0.1, 1.8$\pm$0.2, 1.7$\pm$0.1 &--\\
57.2 & 109.379530 & 37.745110 & -- & -- &--\\
58.1 & 109.379420 & 37.762383 & -- & 4.2$\pm$1.4, 4.2$\pm$1.2, uncons. &3.03$\pm$0.34 \\
58.2 & 109.379260 & 37.762129 & -- & -- & --\\
59.1 & 109.376840 & 37.743245 & -- & 3.4$\pm$0.8, 4.2$\pm$0.5, 3.6$\pm$0.5 &3.11$\pm$0.23\\
59.2 & 109.379960 & 37.748146 & -- & -- &--\\
59.3* & 109.400700 & 37.768628 & -- & -- &--\\
62.1 & 109.372220 & 37.747464 & -- & 2.7$\pm$0.3, 2.9$\pm$0.3, 2.7$\pm$0.3 &2.23$\pm$0.08 \\
62.2 & 109.375510 & 37.752276 & -- & -- & 2.23$\pm$0.06\\ 
63.1 & 109.369420 & 37.756278 & -- & 1.3$\pm$0.1, 1.6$\pm$0.1, 1.7$\pm$0.1 & 2.20$\pm$0.24\\
63.2 & 109.368360 & 37.753695 & -- & -- & 2.20$\pm$0.18\\ 
64.1* & 109.372250 & 37.765391 & -- & -- &--\\
64.2* & 109.371790 & 37.765066 & -- & -- &--\\
\hline
\smallskip
\end{tabular}
\end{center}\caption{Multiple-image systems found after the FF observations.
Images with * are the ones we propose as candidates and that are not used as
constraints in the mass model.
Uncertainties quoted for redshifts predicted by our model, $z_{\rm model}$,  correspond to the 1$\sigma$
confidence level. They are given for the \emph{non-cored}, the \emph{cored}, and the NFW model.
uncons means unconstrained, \emph{i.e.} when the output PDF is flat.
In the last column, we show the estimate of the photometric redshift, when possible, together
with the 1$\sigma$ error bars.
}
\label{multipletable2}
\end{table*}

\begin{table*}
\begin{center}
\begin{tabular}{cccccc}
\hline
ID & R.A. & Dec. & $z_{\rm spec}$ & $z_{\rm model}$ (\emph{non-cored}, \emph{cored}, NFW) & $z_{\rm phot}$\\
\hline \\*[-1mm]

65.1 & 109.383120 & 37.762595 & -- & 5.4$\pm$0.3, $>$4.6, $>$5.0 & --\\
65.2 & 109.382250 & 37.760394 & -- & -- & -- \\
66.1 & 109.383310 & 37.765079 & -- & $>$6.5, 6.4$\pm$0.8, 6.5$\pm$0.4 & --\\
66.2 & 109.380290 & 37.759228 & -- & -- & --\\
67.1 & 109.367470 & 37.756509 & -- & 3.0$\pm$0.1, 4.2$\pm$0.9, 4.5$\pm$0.5 & --\\
67.2 & 109.365120 & 37.749544 & -- & -- & --\\
67.3* & 109.382090 & 37.771629 & -- & -- & --\\
68.1 & 109.392330 & 37.738083 & 5.51 & -- & --\\
68.2 & 109.382350 & 37.736508 & -- & -- & --\\
68.3* & 109.406780 & 37.763803 & -- & -- & --\\
69.1 & 109.380400 & 37.749467 & -- & 3.4$\pm$0.5, 3.3$\pm$0.4, 3.1$\pm$0.4 & --\\
69.2 & 109.375320 & 37.743383 & -- & -- & --\\
69.3* & 109.396440 & 37.768591 & -- & -- & --\\
70.1 & 109.392210 & 37.760496 & -- & 3.7$\pm$0.4, 3.1$\pm$0.2, 3.9$\pm$0.3 & 2.72$\pm$0.06\\
70.2 & 109.389380 & 37.757279 & -- & -- &3.11$\pm$0.07\\
70.3* & 109.375110 & 37.735945 & -- & -- &2.14$\pm$0.09\\
71.1 & 109.382700 & 37.744711 & -- & 3.0$\pm$0.1, 2.6$\pm$0.1, 2.6$\pm$0.1 &2.91$\pm$0.10\\
71.2 & 109.380650 & 37.741346 & -- & -- &1.82$\pm$0.04\\
71.3 & 109.402550 & 37.763610 & -- & -- &2.65$\pm$0.16\\
71.4 & 109.387210 & 37.741295 & -- & -- & -- \\
72.1 & 109.380120 & 37.765496 & -- & $>$4.1, $>$4.0, 3.9$\pm$0.4 &3.32$\pm$0.14\\
72.2 & 109.374810 & 37.762346 & -- & -- &3.32$\pm$0.22\\
73.1* & 109.375970 & 37.748883 & -- & -- &--\\
73.2* & 109.375500 & 37.748483  & -- & -- &--\\
73.3* & 109.391470 & 37.767033  & -- & -- &--\\
74.1* & 109.370890 & 37.751182  & -- & -- & 1.52$\pm$0.06\\
74.2* & 109.383420 & 37.769708  & -- & -- &--\\
74.3* & 109.371310 & 37.751974  & -- & -- &1.60$\pm$0.15\\
75.1 & 109.382000 & 37.738429  & -- & 3.6$\pm$0.4, 4.7$\pm$0.4, 4.3$\pm$0.5  &--\\
75.2 & 109.388640 & 37.739263  & -- & -- &--\\
75.3* & 109.40496 & 37.764678  & -- & -- &--\\
76.1 & 109.411230 & 37.731990  & -- & 4.5$\pm$0.4, 5.4$\pm$0.3,3.7$\pm$0.4 &--\\
76.2 & 109.412840 & 37.733789 & -- & -- &--\\
76.3 & 109.413660 & 37.734646 & -- & -- &--\\
77.1* & 109.377000 & 37.736456 & -- & -- &--\\
77.2* & 109.386260 & 37.751928 & -- & -- &--\\
77.3* & 109.399110 & 37.764959 & -- & -- &--\\
78.1* & 109.371840 & 37.742478 & -- & -- &--\\
78.2* & 109.379010 & 37.752778 & -- & -- &--\\
79.1 & 109.393150 & 37.762795  & -- & 3.7$\pm$0.3, 3.4$\pm$0.2,3.7$\pm$0.3 &--\\
79.2 & 109.387350 & 37.755963 & -- & -- &--\\
79.3 & 109.375740 & 37.735628 & -- & -- &--\\
80.1 & 109.390940 & 37.733628 & -- & 3.7$\pm$0.3, 3.3$\pm$0.1, 3.8$\pm$0.2 & 2.37$\pm$0.15 \\
80.2 & 109.400250 & 37.738496 & -- & -- & 2.50$\pm$0.11 \\
80.3 & 109.410080 & 37.753576 & -- & -- & 2.40$\pm$0.14 \\
81.1* & 109.408320 & 37.728127  & -- & -- & --  \\
81.2* & 109.409010 & 37.728544  & -- & -- & -- \\
81.3* & 109.410160 & 37.729344 & -- & -- & -- \\
82.1* & 109.383660 & 37.766146 & -- & -- & 2.65$\pm$0.48 \\
82.2* & 109.379500 & 37.756303 & -- & -- & 2.72$\pm$0.21 \\
\hline
\smallskip
\end{tabular}
\end{center}
\caption{... continued ... }
\label{multipletable2}
\end{table*}

\end{document}